\documentclass[twocolumn,prb,showpacs]{revtex4}
\usepackage{graphics}
\begin{document}
\title{Local-density approximation for confined bosons in an 
  optical lattice}
\author{Sara Bergkvist}
\email{sara@theophys.kth.se}
\author{Patrik Henelius}
\author{Anders Rosengren}
\affiliation{Condensed Matter Theory, Physics Department, KTH, AlbaNova
  University Center, SE-106 91 Stockholm, Sweden}
\date{\today}

\begin{abstract}
We investigate local and global properties of the one-dimensional
Bose-Hubbard model with an external confining potential, describing an
atomic condensate in an optical lattice. Using quantum Monte Carlo
techniques we demonstrate that a local-density approximation, which
relates the unconfined and the confined model, yields quantitatively
correct results in most of the interesting parameter range. We also
examine claims of universal behavior in the confined system, and
demonstrate the origin of a previously calculated fine structure in
the experimentally accessible momentum distribution.
\end{abstract}

\maketitle

\section{Introduction} 

When a system of cold atoms confined in an external trap is exposed to
a standing-wave laser field the electric field couples to the dipole
moments of the atoms and a so-called optical lattice it created. A
system of trapped bosonic atoms in such a lattice is expected to be
well described by a Bose-Hubbard model.\cite{JaPRL98} The phase
diagram of the homogeneous Bose-Hubbard model was obtained by Fisher
\emph{et al.},\cite{FiPRB89} and confirmed by quantum Monte Carlo
investigations.\cite{BaPRL90} The model exhibits two possible phases
for large on-site interaction between the atoms. At commensurate
filling there is a Mott insulating phase and at non-commensurate
fillings there is a superfluid phase. The optical lattices constitute
a new realization of the Hubbard model, and offer an unprecedented
control of most model parameters such as lattice size, dimension,
chemical potential, and the intersite coupling. Using this high degree
of control Greiner \emph{et al.} have recently studied the release of
atoms from an optical lattice.\cite{GrNat02} In an absorption
measurement they studied the momentum distribution and detected a
gradual decrease of the center peak in the absorption spectra with an
increase in the on-site interaction between atoms, indicating a
growing Mott insulating phase in the center of the trap. The need to
interpret and understand this, and related, recent experiments
performed in optical lattices have renewed interest in understanding
various extensions of the well-studied bosonic Hubbard model.

In the case of optical lattices the picture is complicated by the
trapping potential. The trapping potential is an external potential
applied to keep the atoms in the lattice. Due to the spatial variation
of the confining potential, different phases can be realized in
different parts of the trap, leading to a spatial phase separation.
This has been demonstrated using quantum Monte Carlo techniques for
bosons\cite{BaPRL02} and fermions.\cite{RiPRL03, Ri03} These studies
focused on the differences between the phase diagram of the confined
and unconfined model and calculated state-diagrams for the confined
model. A density matrix renormalization group study\cite{KoPRA04}
shows that the decay of correlation functions in the confined case can
be easily rescaled to obtain the form expected for the homogeneous
model. These results were also shown to agree very well with a
hydrodynamical treatment of the 1D Bose gas combined with a local
density approach. Results for large systems have also been obtained by
combining the Gutzwiller mean-field ansatz with a numerical
renormalization group procedure.\cite{PoPRA04} Recent large-scale
Monte Carlo studies\cite{We04a,We04b} extend previous work to higher
dimensions and address questions of critical behavior and local order
parameters.
 
In this work we focus on the similarities of the phase diagram for the
unconfined and the confined state diagram. We map the site-dependent
confining potential for the trapped system to the chemical potential
for the homogeneous system and in this manner determine the properties
of the confined system in a local-density approximation. It is
reasonable to believe that a good agreement between the confined and
the unconfined system is obtained for local observables in weakly
interacting systems, where the correlation length is short. Here we
demonstrate that the approximation can be used with good accuracy in
most of the range of interactions where the Mott insulating phases
appear. Furthermore, the approximation can be used also for the
nonlocal momentum distribution, which is calculated from the
particle-particle correlation function. This local-density
approximation has been used to compare a number of observables
previously,\cite{BaPRL02,RiPRL03,Ri03,PoPRA04,We04a,We04b} but here we
present a more exhaustive comparison of several observables. One
practical aspect of the local-density approximation is that it allows
a very quick estimate of the distribution of different phases in a
specific trap setting from knowledge of only the phase diagram of the
homogeneous model.

The outline of the paper is as follows. In Sec.~\ref{SecBos} we
discuss the phase diagram for the Bose-Hubbard model and explain
the mapping between the confined and the unconfined model. In
Sec.~\ref{SecRes} we briefly discuss the Monte Carlo method and
present results testing the validity of the local-density
approximation. We conclude with a summary and conclusions.

\section{Bosons in an optical lattice}
\label{SecBos}
\begin{figure}
\begin{center}
 \resizebox{60mm}{!}{\includegraphics{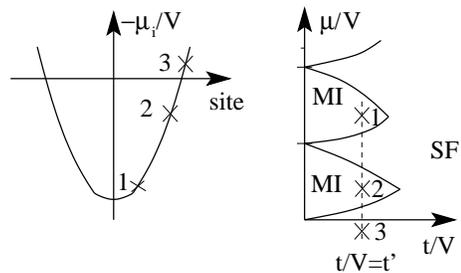}}
\end{center}
\caption{Demonstration of the local-density approximation where a
  system with a confining potential is mapped to a slice in the phase
  diagram for the Bose-Hubbard model. The figure to the left
  represents the site dependent potential $\mu_i/V$ as a function of
  lattice site for a confined system with hopping parameter
  $t/V=t'$. The figure to the right shows a schematic phase diagram
  for the unconfined Bose-Hubbard model, where the insulating Mott
  lobes (MI) and superfluid phase (SF) are indicated. The dashed line
  in the phase diagram represents the different values of the
  effective chemical potential obtained in the confined system. The
  three x's mark how three specific sites in the confined system are
  mapped to the phase diagram on the right.}
\label{Figslice}
\end{figure}

Starting from a general Hamilton operator for bosonic atoms in an
external trapping potential and optical lattice, Jaksch \emph{et
al.}\cite{JaPRL98} expand the bosonic field operators in the Wannier
basis and arrive at the Bose-Hubbard model. In order to review the
properties of this fundamental model we start by disregarding the
effects of the confining potential and consider the homogeneous
one-dimensional Bose-Hubbard model,
\begin{equation}
H_{BH}=\sum_{i} -t(c^{\dagger}_i
c_{i+1}+\hbox{H.c.})+\frac{V}{2}n_i(n_i-1)-\mu n_i,
\end{equation}
where $t$ is the hopping matrix element between adjacent sites, $V$ is
the on-site repulsion, $\mu$ is the chemical potential, and $n_i$ is
the number of bosons at site $i$. At zero temperature there are two
possible phases for the model. For small on-site repulsion the kinetic
energy is minimized by a phase coherent superfluid phase,
characterized by a nonzero superfluid density, algebraically decaying
correlation functions and a linear, gapless excitation spectrum. As
the on-site repulsion is increased it becomes harder for the bosons to
move, due to the increasing potential energy cost. At a critical value
of the on-site repulsion the model enters the Mott insulating phase
where each lattice site is filled with the same number of bosons. The
suppression of the fluctuations in particle number makes the Mott
insulating phase incompressible with a vanishing compressibility,
$\kappa=\beta(\langle n^2\rangle -\langle n\rangle^2)=0$. The
insulating phase is gapped and correlation functions decay
exponentially. The zero-temperature phase diagram for the Bose-Hubbard
model, as a function of density and the hopping parameter $t/V$, is
schematically shown in the right part of Fig.~\ref{Figslice}. For
small values of $t/V$ there is a series of Mott lobes as the density
is increased. In each Mott lobe the system is in the insulating phase
and the density is constant.  At large values of $t/V$ the model is in
the superfluid phase.\cite{FiPRB89}

As shown by Jaksch \emph{et al.}\cite{JaPRL98} the Bose-Hubbard model
is an effective model for a Bose-Einstein condensate of atoms when
subjected to an optical lattice potential. The potential energy is
given by $V=4\pi a_s \hbar \int d^3x|w(x)|^4/m$, where $a_s$ is the
$s$-wave scattering length, $w(x)$ is a localized Wannier function and
$m$ is the atomic mass.  As the lattice potential is turned on the
condensate remains in a superfluid state for small values of the
potential. However, as the potential is increased there is a
transition to an insulating phase where the phase coherence of the
condensate is destroyed, as demonstrated by Greiner \emph{et
al.}.\cite{GrNat02} Other properties of the Mott insulating phase,
such as the robustness to external perturbations due to the excitation
gap have also been verified experimentally.\cite{GrNat02} The
relevance of the Bose-Hubbard model to Bose-Einstein condensates have
therefore been demonstrated both experimentally and theoretically, and
in this work we examine the model further using quantum Monte Carlo
simulations. Some of the quantities we examine, such as the density
profile of bosons in the trap, or the momentum distribution of the
bosons, can be directly measured experimentally. Other quantities,
such as local-density fluctuations may be harder to observe
experimentally, but are of interest since they tell us how the
superfluid-insulator transition occurs in a confined system.

In experimental realizations of the optical lattice a confining
potential is necessary to prevent the bosons from escaping. The
confinement typically spans a couple of hundred lattice sites and its
strength is adjusted to be large enough so that there are no atoms at
its edges. The confining potential is also used to control the
dimensionality of the system. By using a very narrow confinement in
some directions the dimensionality of the system is changed. In this
work we consider a setup with a narrow confinement in two dimensions
and a broad confining potential in one dimension, leaving a
one-dimensional system. The broad confining potential adds a site
dependent potential term to the Hamiltonian
\begin{eqnarray}
H&=&H_{BH}+U_c\sum_{i}\left|i-\frac N2\right|^{\alpha}n_i\\\nonumber
&=&\sum_{i} -t(c^{\dagger}_i
c_{i+1}+\hbox{H.c.})+\frac{V}{2}n_i(n_i-1)-\mu_i n_i,
\label{confpot}
\end{eqnarray}
where $\alpha$ is a parameter that controls the power of the confining
potential. In the second line of the equation the confining potential
and the chemical potential $\mu$ are combined into a site dependent
chemical potential, $\mu_i$. Different sites in the lattice therefore
have different effective chemical potentials. In the limit of
vanishing $t$ the local properties of the system are determined by the
value of the local chemical potential. For finite values of $t/V$ we
assume that the local properties of the confined system may be
obtained from a homogeneous system with the same value of $t/V$ as in
the confined system, and a chemical potential equal to the effective
chemical potential at the point considered in the confined
lattice. This mapping is illustrated in Fig.~\ref{Figslice} where it
is shown how the confined system represents a constant $t/V$ slice in
the $\mu-t$ diagram of the homogeneous system.

We examine the accuracy of this local-density approximation for a
range of hopping parameters where the Mott insulating phases are
present. In general we expect poor results for very deep and narrow
traps where the potential varies rapidly from site to site and close
to phase transitions where the correlation length diverges in the
unconfined system. At a true phase transition the correlation length
diverges, but due to the limited size of the confined system the
correlation length cannot diverge and there cannot be a true phase
transition. We merely expect a system with spatially separated regions
of different phases, a scenario which is supported by our results. A
careful analysis of the lack of critical behavior can also be found
in Refs.~[\onlinecite{We04a,We04b}].

The trap used in the experiment described by Greiner was of size
$65\times 65 \times 65 $ and contained a maximum of 2.5 bosons per
site. In this work we consider one-dimensional traps with a linear
size of about 300 sites and a density up to two bosons per
site. Therefore, we believe that the results we obtain are pertinent to
recent experimental setups.

\section{Results}

In this work we use the stochastic series expansion quantum
Monte Carlo method.\cite{SaPRB99,SyPRE02} We perform the
calculations at a fairly low temperature of $\beta/V=256$, and the
convergence is controlled with a calculation at a temperature of half
this value.

We study the densities in a trap with a linear, square, and a quartic
potential, i.e. $\alpha=$ 1, 2, and 4 in Eq.~(\ref{confpot}). The
calculation is performed for three different values of the hopping
parameter, $t=$ 0.05, 0.1, and 0.2. We mainly show results for the two
cases $t=$ 0.1 and 0.2, since $t=$ 0.05 and $t=0.1$ yield
quantitatively very similar results. The on-site repulsion $V$ is set
to one. The strength of the confining potential $U_c$, and the
chemical potential $\mu$, are adjusted so that the effective chemical
potential $\mu_i/V=1$ in the center of the well, and is large enough to
ensure that there are no bosons at the edges of the trap.

\begin{figure}
\begin{center}
  \resizebox{60mm}{!}{\includegraphics{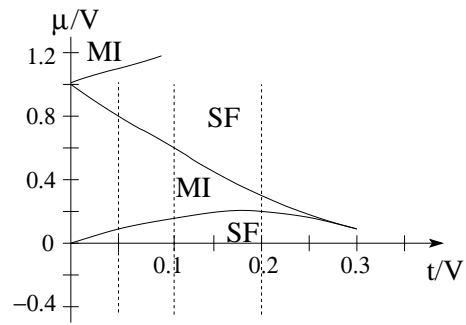}}
\end{center}
\caption{A detailed phase diagram showing the first Mott lobe (MI) and
  the surrounding superfluid phase (SF). The chemical potential $\mu
  /V$ is plotted as a function of the hopping parameter $t/V$. The
  three slices of the $\mu-t$ diagram that our confined systems
  represent are indicated with the dotted lines. The dotted lines
  start at a $\mu$ low enough to ensure that there are no atoms at the
  edges of the trap, and they all end at $\mu/V=1$.}
\label{FigPhaseD}
\end{figure}

In Fig.~\ref{FigPhaseD} we display the approximate phase diagram of
the homogeneous Bose-Hubbard model, based on Fig. 5 in
Ref.~[\onlinecite{BaPRB92}] and Fig. 9 in Ref.~[\onlinecite{KuPRB00}]. In
the phase diagram we indicate the three slices at fixed value of the
hopping $t/V$ that our confined models represent. We note that the
slices at $t/V=0.05$ and $t/V=0.1$ cut through a significant part of
the first Mott lobe, while $t/V=0.2$ is chosen to cut through the tip
of the same lobe.

\begin{figure}
\begin{center}
 \resizebox{70mm}{!}{\includegraphics{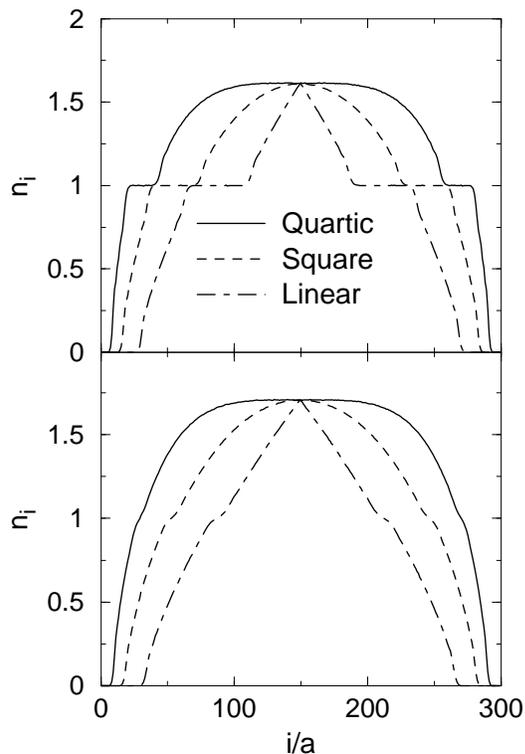}}
\end{center}
\caption{The local-density $n_i$ as function of lattice sites for
three different confining potentials. The upper figure is calculated
for $t/V=0.1$ and the lower for $t/V=0.2$. The lattice
constant is denoted by $a$.}
\label{FigDens}
\end{figure}

The local-density as a function of lattice site is shown in
Fig.~\ref{FigDens}. We note that for $t=0.1$ there is a range of
lattice sites for which the local-density takes the value of one boson
per site for all three confining potentials. Since the density
fluctuations are strongly suppressed in these regions they appear
incompressible and are therefore called Mott plateaus, indicating that
the system is in a Mott insulating state.  For $t=0.2$ there is not
yet a well defined plateau, but all three density curves corresponding
to the three different confining potentials display a slight kink at
unit density.

\label{SecRes}

\begin{figure}
\begin{center}
  \resizebox{70mm}{!}{\includegraphics{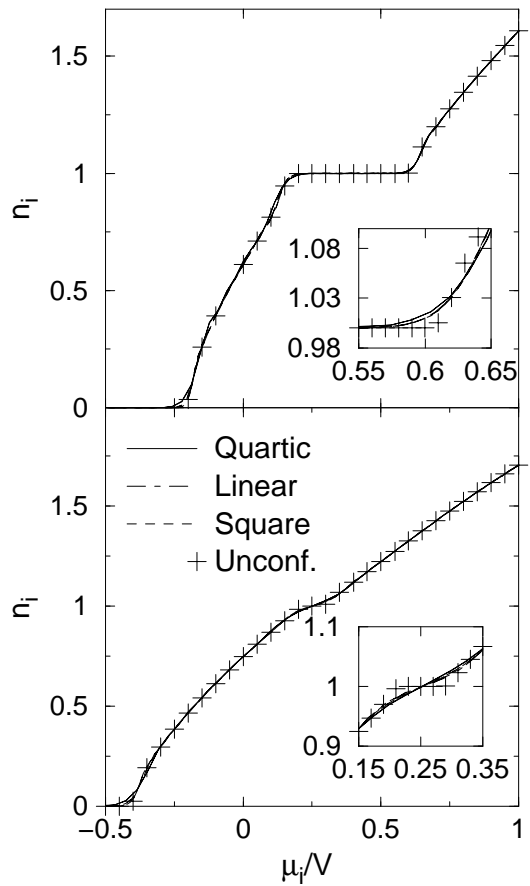}}
\end{center}
\caption{A comparison of the densities for the confined and unconfined
case. The local-density $n_i$ is plotted as a function of the
effective chemical potential $\mu_i/V$. Note that the results for the
quartic, linear, and square potentials essentially coincide with each
other, and with the results for the unconfined system. The insets
show the region around points where there is a phase transition for
the homogeneous system. In the upper figure $t/V=0.1$ and in the lower
figure $t/V=0.2$.}
\label{FigDensV}
\end{figure}

If the densities at different lattice sites are presented as a
function of the effective on-site chemical potential, this density can
be directly compared with the density in a homogeneous system with the
same value of the chemical potential. The results are shown in
Fig.~\ref{FigDensV}, and we see a good quantitative agreement between
the densities for the confined and the unconfined systems for most of
the parameter regime. Due to the diverging correlation function there
is some discrepancy close to the place where there is a
phase separation between the Mott insulating regions and the
superfluid phases, which is shown in the insets. The sharp edge for
the homogeneous system is simply smoothed out, indicating the absence
of a true phase transition as the plateau is approached in the
confined system. In the lower part of the figure, which presents
results for $t/V=0.2$, one can see that there is a small plateau in the
homogeneous case, which is smeared out in the confined systems. This
indicates that there is no true Mott insulating region in the confined
system even though the unconfined system has a Mott insulating
phase. We also can detect a small difference close to the edge of the
trap, where the local-density decreases quite rapidly from site to
site.

\begin{figure}
\begin{center}
  \resizebox{70mm}{!}{\includegraphics{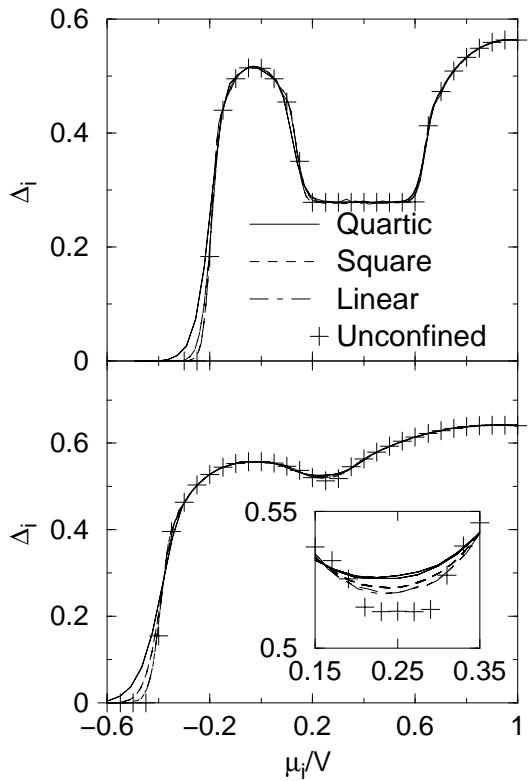}}
\end{center}
\caption{Variance in the local-density $\Delta_i$ as a function of the
effective potential $\mu_i/V$ for different confining potentials and
also for the unconfined system. In the upper figure $t/V=0.1$ and in
the lower figure $t/V=0.2$.}
\label{FigKomp}
\end{figure}

The Mott insulating phase is incompressible, and the global
compressibility vanishes in the Mott insulating phase for the
unconfined system. For the confined system the global compressibility
never vanishes\cite{BaPRL02} since there are always superfluid
regions present. Here we examine the behavior of a local
compressibility, defined as the fluctuations in the local-density
\begin{equation}
\Delta_i^2=\langle n_i^2\rangle-\langle n_i\rangle^2.
\end{equation}
The results are presented in Fig.~\ref{FigKomp} where we once again
compare the result for the confined system with values obtained for
the unconfined case. Again we notice a good general agreement, but
discrepancies close to the edge of the plateaus and particularly at
the edge of the trap. The quartic potential rises faster than the
other potentials at the boundary of the trap, and it is also in the
data for the quartic potential that we see the largest deviations from
the homogeneous case. For $t=0.2$ the inset shows the deviations in
the region where the homogeneous system forms a Mott plateau. This is
another indication that the confined system does not manage to form a
well developed insulator here, as mentioned above in the argumentation
about the density profile. This has previously been noted for the case
of fermionic systems.\cite{Ri03}

\begin{figure}
\begin{center}
  \resizebox{70mm}{!}{\includegraphics{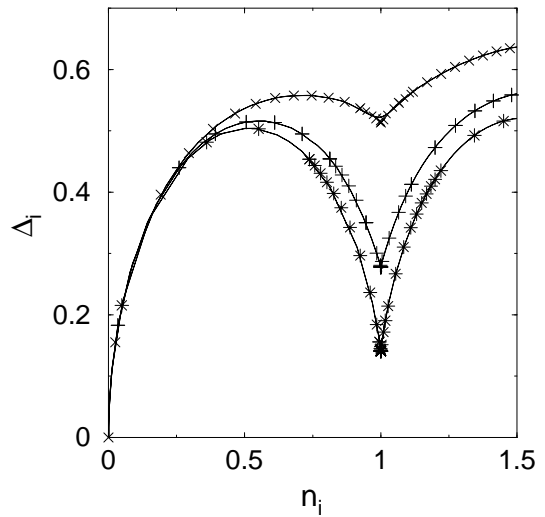}}
\end{center}
\caption{Variance in the local-density $\Delta_i$ as a function of
density for the unconfined case [$t/V=0.05$ ($\ast$), $t/V=0.1$ (+),
$t/V=0.2$ ($\times$)]. The solid line shows results for the
confined system with a square potential.}
\label{FigKompDens}
\end{figure}

Next we focus on the properties of the system as a plateau is
approached. As a Mott plateau is entered by scanning the chemical
potential for the homogeneous system there is a second-order phase
transition from the superfluid to the Mott insulating phase. In the
confined system, at least in one dimension, there can be no true phase
transition since there cannot be a divergent correlation length at the
boundary between the two phases, and we simply have a system
displaying spatial phase separation. A similar conclusion was also
reached for a two-dimensional system.\cite{We04a,We04b} Nevertheless, there
have been claims of universal behavior as a Mott plateau is
approached.\cite{BaPRL02,RiPRL03} In Ref.~[\onlinecite{BaPRL02}] it was
argued that the variance in the local-density shows universal behavior
as the Mott plateau is approached. This behavior was not reproduced in
a numerical renormalization group study\cite{PoPRA04}, and in
Fig.~\ref{FigKompDens} we display the variance in the local-density as
the Mott lobe is approached for three different values of the hopping
parameter. By analyzing the data in Fig.~\ref{FigKompDens} more
closely we find that the local-density approximation works very well,
in contrast to the fermionic case.\cite{RiPRL03} The accuracy of the
local-density approximation probably explains the ``universal''
behavior seen previously. However, as $t$ is varied the curves
readily separate and there is thus no real universality.

Another case of universality as the Mott plateau is approached in
a confined system has been reported\cite{RiPRL03} for a different
definition of a local compressibility,
\begin{equation}
\kappa_i^l=\sum_{|j|\le l}\langle n_in_{i+j}\rangle -\langle
n_i\rangle\langle n_{i+j}\rangle,
\label{localcorr}
\end{equation}
which reflects the response at site $i$ to a change in the chemical
potential in a region of size $l$. This compressibility was introduced
as a local order parameter, which vanishes in the insulating phase. It
was also argued that this compressibility decays with a universal
power law as a Mott plateau is approached in a fermionic system. We
investigate this quantity for a bosonic system. The value of $l$ is
somewhat arbitrary and was previously chosen to be larger than the
correlation length in the insulating phase. Here we take the sum over
the entire system, to remove the uncertainty associated with the size
of $l$, and hereafter we remove the index $l$. The compressibility
still behaves in essentially the same way as if we limit the sum to
some smaller region. 

Extending the range of the summation makes the definition of the local
compressibility identical to the definition used in
Refs.~[\onlinecite{We04a,We04b}], and the sum of the local
compressibility at all sites is the global compressibility. This
compressibility diverges for the homogeneous system as the phase
transition is approached from the superfluid phase, and is zero in the
insulating phase, as mentioned above. Finite-size effects prevent the
compressibility from diverging in a finite system, and instead the
compressibility approaches zero as a smooth function as the insulating
state is entered. The way in which the local compressibility
approaches zero in a finite but unconfined system is therefore a
finite-size effect.

\begin{figure}
\begin{center}
 \resizebox{70mm}{!}{\includegraphics{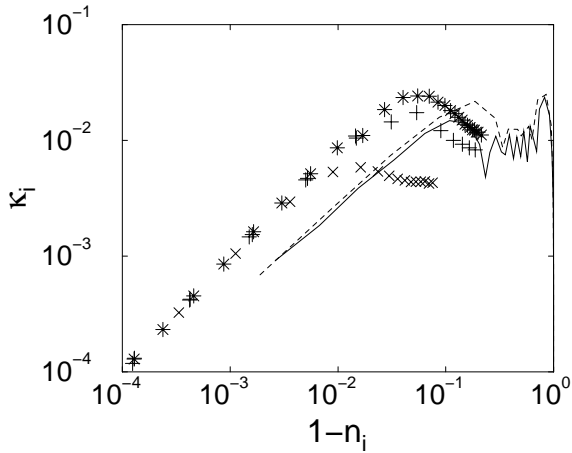}}
 \end{center}
\caption{Local compressibility $\kappa_i$ for the unconfined case
 [$t/V=0.05$ ($\ast$), $t/V=0.1$ (+), $t/V=0.2$ ($\times$)] and the
 confined system with a square potential [$t/V=0.05$ (dashed line),
 $t/V=0.1$ (solid line)].}
\label{FigDensCorr}
\end{figure}

In Fig.~\ref{FigDensCorr} we show this decay of the local
compressibility as the Mott insulator is approached. It appears that
the compressibility decays algebraically as we approach the Mott
plateau, which is the behavior described for the fermionic
system\cite{RiPRL03}. The exponent is very similar for the different
strengths of the hopping. However, we also note that the
compressibility for the homogeneous system follows the same behavior
as the confined system, in contrast to the behavior depicted for the
fermionic case. It appears that this algebraic decay is not the result
of universal behavior as a Mott plateau is approached in the
confined system, but rather another property of the homogeneous model
that carries over to the confined system. It is reasonable that the
confined and the unconfined model display the same type of finite-size
effects, and therefore the observed similarity between the confined and
the unconfined system is expected. In the bosonic case we find that
the exponent that describes the decay has a numerical value of about
1.0, as compared to about 0.7-0.8 in the fermionic case.\cite{RiPRL03}
The exponent appears slightly smaller for the confined systems, but
the statistical fluctuations are larger in this case, which probably
explains the small difference. The same exponent is obtained if we
analyze the transition to zero density at the edge of the trap, which
is expected since this is the same kind of phase transition.

We have done a short calculation for the fermionic system and we found
that for both the unconfined and the confined systems the local
compressibility decays to zero in an algebraic manner. This is
different from the previously reported behavior\cite{RiPRL03}, described above,
where only the confined system displays such a decay. A possible
reason for the discrepancy is that we use a grand canonical ensemble,
instead a of canonical ensemble, which seems to be more appropriate
when observing particle number fluctuations. We found that the exponent
for the unconfined model is very close to one and that the confined
model may have the same, or a slightly, smaller exponent.

By releasing the atoms in the trap the momentum distribution, $n_k$,
can be obtained experimentally.\cite{GrNat02} This is probably the
most accessible observable that can be studied both experimentally and
theoretically. The momentum distribution is the Fourier transform of
the particle-particle correlation function,
\begin{equation}
n_k=\sum_{i,j}\exp^{i k(r_i-r_j)}\langle c_i^{\dagger}c_j\rangle .
\end{equation} 
In the homogeneous model the particle-particle correlation function is
expected to decay algebraically in the superfluid phase, while it
decays exponentially in the insulating phase. By a simple rescaling
this has been demonstrated also for the trapped system.\cite{KoPRA04}

\begin{figure}
\begin{center}
 \resizebox{70mm}{!}{\includegraphics{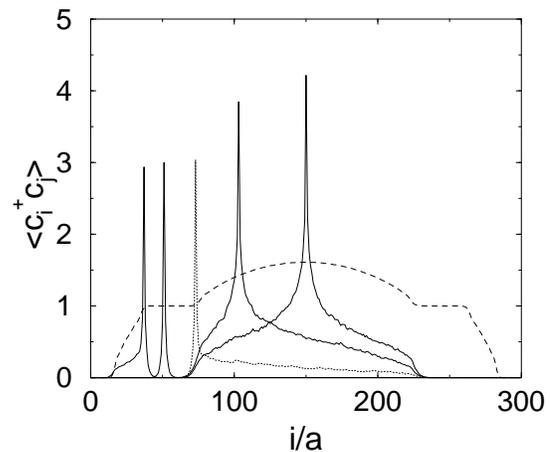}}
 \end{center}
\caption{The particle-particle correlation function $\langle
 c_i^{\dagger}c_j\rangle$ for a few selected lattice points $i$ to all other
 points $j$ in the lattice. The dashed line represents the density
 profile, shown here to indicate the location of the Mott insulating
 phases. The dotted line marks the correlation function from a point
 on the phase separation boundary between the super fluid and Mott
 insulating regions.}
\label{FigCorr}
\end{figure}

In Fig.~\ref{FigCorr} we show the particle-particle correlation
function from a few selected points in the trapped system. We can
clearly see the exponential decay of the correlation function within
the insulating phase, and the much slower decay in the superfluid
phase. We also note that correlation functions from points in the
superfluid regions decay exponentially as they enter the Mott
insulating regions. Especially interesting are points close to the
phase separation boundary. The dotted line represents the correlation
function from such a point, which displays exponential decay on one
side and a long-range algebraic decay on the other side.

\begin{figure}
\begin{center}
  \resizebox{70mm}{!}{\includegraphics{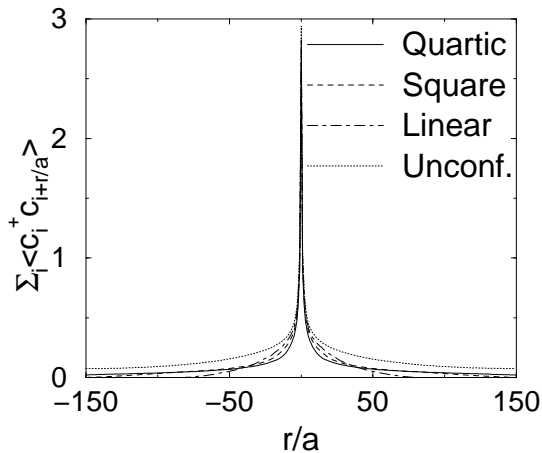}}
 \end{center}
\caption{The spatial average of the particle-particle correlation
function for the confined system is compared with the value for the
homogeneous system for $t/V=0.1$.}
\label{FigCorrSum}
\end{figure}

To obtain the momentum distribution we first average the
particle-particle correlation function over the whole system. In
Fig.~\ref{FigCorrSum} the sum of all the particle-particle correlation
functions is presented. This function is compared with the function
obtained if the correlation functions for different $\mu$-values for
the homogeneous system are added. The chemical potentials are chosen
with equal spacing in the whole range covered by the effective
potential in the well. The sum can immediately be compared with the
total correlation function for the system with a linear confining
potential. However, for the other potentials the correlation functions
have to be reweighted since we use equally spaced values of the
chemical potential in the homogeneous system. To reweight the
correlation function we multiply the distribution with the derivative
 of the confining potential with respect to $i$, the
lattice position. We notice that the agreement is very good for small
distances, while the correlation function for the homogeneous system
has a longer tail, which can be expected, since the insulating phases
cut off the correlations in the trapped system.

\begin{figure}
\begin{center}
  \resizebox{70mm}{!}{\includegraphics{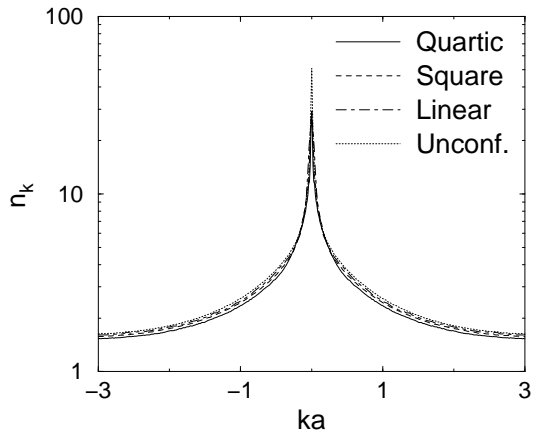}}
 \end{center}
\caption{The momentum distribution for the confined system is
compared with the value for the homogeneous system for $t/V=0.1$.}
\label{FigMom}
\end{figure}

When the correlation function is Fourier transformed to get the
momentum distribution, the difference at large distances in the
correlation function gives a difference in the momentum distributions
for small $k$ values. The momentum distribution for the confined
systems is compared with the momentum distribution for the homogeneous
system in Fig.~\ref{FigMom}, and except for the peak value we notice
a reasonable overall agreement.

\begin{figure}
\begin{center}
  \resizebox{70mm}{!}{\includegraphics{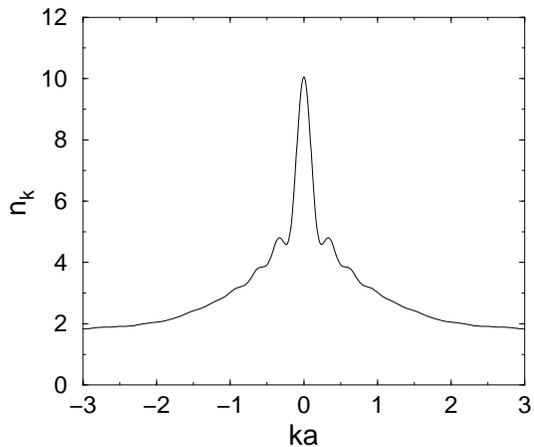}}
 \end{center}
\caption{An example of the fine structure when the particle-particle
correlation function from one point in the superfluid phase is Fourier
transformed.}
\label{FigFineStr}
\end{figure}

Previous calculations have indicated that the appearance of a fine
structure in the momentum distribution can be used to detect the
formation of a Mott insulating plateau in the middle of the
well.\cite{KaPRA02} In another study the fine structure is claimed to
be a finite-size effect.\cite{PoPRA04} We observe fine structure in
the Fourier transform of correlation functions originating from
individual sites in the superfluid phase. This structure appears
as a consequence of the abrupt cut in the correlation function
obtained when the Mott insulating phases or the edges of the system
are reached, as mentioned above. An example of this structure is
shown in Fig.~\ref{FigFineStr}. However, the sum of all the
correlations is smooth since the cutoff in the correlation functions
appears at different distances for different lattice points. In case a
small lattice is studied some fine structure may remain and this
supports the earlier conclusion\cite{PoPRA04} that the satellite peaks
are a finite-size effect.

\section{Conclusions}

We examine the accuracy of a local-density approximation for the
one-dimensional Bose-Hubbard model using a quantum Monte Carlo method.
In this approximation the behavior of the confined system is
approximated with that of the unconfined system. In most of the
interesting parameter regime one can obtain quantitatively accurate
results using this approximation. In this manner the state diagram for
a confined system can be determined to a high degree of accuracy from
knowledge of the phase diagram of the homogeneous Bose-Hubbard model.
We demonstrate how the approximation fails where the correlation
length diverges in the homogeneous system, and also when the local
density varies rapidly from site to site. Even for the case of the
nonlocal momentum distribution the local-density approximation works
fairly well. Contrary to earlier studies we find no evidence of
universal scaling as an insulating plateau is approached. Since the
confined systems do not display a continuous phase transition this is
to be expected. Finally, we provide more evidence for the case that an
observed fine structure in the momentum distribution is related to a
finite-size effect.

\begin{acknowledgments}
We are indebted to S. Wessel for instructive discussions on the local
 compressibility and to A. Sandvik and O. Sylju{\aa}sen for fruitful
 discussions. This work was supported by the Swedish Research Council
 and the G\"oran Gustafsson foundation.
\end{acknowledgments}

\bibliography{bib} \end{document}